\documentclass[11pt,a4paper]{article}
\pdfoutput=1
\usepackage{jcappub}
\usepackage{float}
\usepackage{graphicx}
\usepackage{subfigure}
\usepackage{epsfig}
\usepackage{epstopdf}
\usepackage{indentfirst}
\newcommand{\be}{\begin{equation}}
\newcommand{\ee}{\end{equation}}

\newcommand{\ba}{\begin{eqnarray}}
\newcommand{\ea}{\end{eqnarray}}

\begin{document}

\title{A de Sitter tachyonic braneworld revisited}
\author[a]{Nandinii Barbosa--Cendejas}
\author[b]{Roberto Cartas--Fuentevilla}
\author[b]{Alfredo Herrera--Aguilar}
\author[c]{Refugio Rigel Mora--Luna}
\author[d]{Rold\~ao da Rocha}

\affiliation[a] {Facultad de Ingenier\'\i a
El\'ectrica, Universidad Michoacana de San Nicol\'as de Hidalgo.
Edificio $\Omega$, Ciudad Universitaria, C.P. 58040, Morelia,
Michoac\'{a}n, M\'{e}xico.}
\affiliation[b]{Instituto de F\'{\i}sica, Benem\'erita Universidad
Aut\'onoma de Puebla. Apdo. Postal J-48, C.P. 72570, Puebla, Puebla ,
M\'{e}xico.}
%\affiliation[c]{Instituto de F\'{\i}sica y Matem\'{a}ticas,
%Universidad Michoacana de San Nicol\'as de Hidalgo.
%Edificio C--3, Ciudad Universitaria, C.P. 58040, Morelia, Michoac\'{a}n, M\'{e}xico.}
\affiliation[c] {Facultad de Ingeniera Qu\'imica,
Universidad Michoacana de San Nicol\'as de Hidalgo.
Edificio $M$, Ciudad Universitaria, C.P. 58040, Morelia,
Michoac\'{a}n, M\'{e}xico.}
\affiliation[d] {Centro de Matem\'atica, Computa\c c\~ao e Cogni\c c\~ao,
Universidade Federal do ABC\\
09210-580, Santo Andr\'e, Brazil}

\emailAdd{rcartas@ifuap.buap.mx}
\emailAdd{nandinii@ifm.umich.mx}
\emailAdd{aherrera@ifuap.buap.mx}
\emailAdd{rigel@ifm.umich.mx}
\emailAdd{roldao.rocha@ufabc.edu.br}

\abstract{Within the framework of braneworlds, several interesting
physical effects can be described in a wide range of energy scales,
starting from high-energy physics to cosmology and low-energy
physics.
%Braneworld effects modify standard 4D results and attempt
%to solve some open problems of modern physics, moreover, they can be
%tested by making use of highly precise experiments and/or observational
%data which, in turn, set bounds on the braneworld model parameters.
An usual way to generate a
thick  braneworld model relies in coupling a bulk scalar field to
higher dimensional warped gravity. Quite recently, a novel
braneworld was generated with the aid of a tachyonic bulk scalar
field, having several remarkable properties. It comprises a  regular and stable solution that contains a relevant 3--brane with
de Sitter induced metric, arising as an exact solution to the 5D
field equations, describing the inflationary eras of our Universe. Besides, it is {\it asymptotically flat}, despite of the presence of
a negative 5D cosmological constant, which is an interesting feature that
contrasts with most of the known, asymptotically either dS
or AdS models. Moreover, it encompasses a graviton spectrum with a single
massless bound state, accounting for 4D gravity localized on the
brane, separated from the continuum of Kaluza-Klein massive graviton
modes by a mass gap that makes the 5D corrections to
Newton's law to decay exponentially. Finally, gauge, scalar and fermion
fields are also shown to be localized on this braneworld. In this
work, we show that this tachyonic braneworld allows for a nontrivial
solution with a vanishing 5D cosmological constant that preserves
all the above mentioned remarkable properties with a less amount of
parameters, constituting an important contribution to the
construction of a realistic cosmological  braneworld model. }

\maketitle

%\arxivnumber{arXiv:***}
\flushbottom

%\pacs{11.25.-w, 04.50.-h, 04.50.Gh}

\section{Introduction}

In the last two decades, the paradigm underlying the observable Universe as a 3--brane, embedded in a higher-dimensional spacetime, has  become a fruitful
scenario for addressing several questions in physics,  that cover a comprehensive plethora of phenomena, ranging from low energy physics \cite{lep1,lep2,lep3,lep4}, gravity \cite{gravity1,gravity2,gravity3,Bazeia:2004dh,Bazeia:2008zx},
astrophysics \cite{astrophysics1,astrophysics2,astrophysics3,astrophysics4}, and high energy physics \cite{hep1,hep2,hep3,hep4,hep5} to cosmology
\cite{cosmology1,cosmology2,cosmology3,cosmology4,cosmology5}. Regarding some prominent cases, the braneworld paradigm leads to plausible
reformulations or even complete solutions of these problems (see  Refs.
\cite{Liu:2017gcn,Dutra:2014xla,thbrs,Maartens:2010ar} for complete reviews). A striking property of these models consists of pointing out that a higher dimensional world could encode our universe, without any conflict with 4D current experiments and
recent observational data.

In the context of the braneworld physics, there is a branch of models in
which the fifth dimension is modeled by bulk scalar fields, extending the idea of
thin branes to thick brane configurations. Several different thick
brane configurations, whose dynamics is generated by a 5D gravitational
action with a bulk scalar field and a self-interacting potential have been proposed. The nature of the scalar field setup leads to different specific scenarios (see the
reviews \cite{Liu:2017gcn,Dutra:2014xla,thbrs}).

Despite the usefulness of the braneworld paradigm for approaching the above
mentioned problems,  the quest for  a more realistic approach imposes
a series of important tests and physical constrains.  Namely,  the
braneworld models, whose signatures must be reflected as small corrections to the 4D
physical laws according  to experimental and observational data,  must be able to recover the 4D physics of our
Universe for precise physical limits. Hence, it is essential to define a robust physical notion of 4D gravity
and matter fields localization on the brane within these models with the less possible amount of parameters.

In order to get a consistent physical model the braneworld configuration must be
{\it stable} under small fluctuations of all the gauge, matter, and gravity background fields involved in the setup
 \cite{KKS}. Checking for this kind of stability is a highly
non-trivial task, strongly dependent on the field content of the model. Another
important aspect is the localization of the Standard Model matter (gauge, scalar and
fermion fields) onto the brane \cite{fieldslocalization1,fieldslocalization2,fieldslocalization3,fieldslocalization4,fieldslocalization5,fieldslocalization6,fieldslocalization7,alencar1}. So far,
different scenarios of thick branes have been studied in the literature, revealing that the localization mechanism of matter fields directly depends on the warp factor into the 5D metric \cite{thickbraneworlds}.

Within this braneworld framework, a tachyonic scalar field has been used to generate
such models with recent applications to cosmology \cite{tachyoncosmology,mazumdar},
localization of 4D gravity on expanding 3--branes \cite{GADRR,tachyonbrane1,tachyonbrane2} and
localization of the Standard Model matter fields \cite{coulombcorrec,ahagauge,AARRH,fieldslocalizationtachyon}.

The original form of the tachyonic effective action was proposed along a series of articles \cite{Sen0,garousi,bergshoeff,kluson} as a supersymmetric generalization of the Dirac--Born--Infeld action that describes the dynamics of light modes (massless and tachyonic) on the world-volume of a non-BPS D-brane within the context of type II string theory in Minkowski spacetime \cite{senWSL1,senWSL2}.
%It turns out that D-branes can give rise to both stable BPS states and unstable objects such as brane-antibrane configurations and non-BPS D-branes, the latter are unstable and can decay to stable D-branes. Moreover, the non-BPS Dp-branes in these theories are related to BPS D(p+1)-brane-antibrane configurations by condensation of the tachyon living on this brane-antibrane pair. Thus, the tachyonic effective field theory describing the dynamics of a non-BPS D-brane in string theory possesses a BPS D-brane. By studying the world-volume theory of massless modes on this BPS D-brane, it was shown that the world volume action has precisely the Dirac-Born-Infeld form without any higher derivative corrections \cite{sen4}.
%
%\footnote{The tachyon field we consider in the action (\ref{ta}) is real in contrast to the fact that the original string theory effective action involves a complex tachyon field that lives in the brane-antibrane configuration.}
%
This tachyonic effective action has found several applications within string cosmology \cite{stringcosmo1,stringcosmo2,stringcosmo3,stringcosmo4,stringcosmo5,stringcosmo6} and  supergravity \cite{bazeia}. Moreover, solar system constraints were imposed on its parameters by considering this action as a scalar-tensor model \cite{devi}.

From the cosmological point of view, the early Universe and the
observed accelerating expansion can be qualitatively described
within this de Sitter braneworld model. Thus, these attempts to
construct a consistent inflationary braneworld take into account the
fact that the cosmic inflationary theory is in good agreement with
temperature fluctuation properties of the Cosmic Microwave
Background Radiation, and that the inflationary epoch likely took
place at very high temperatures \cite{tachyoniflation}.

Thus, the construction of a tachyonic braneworld model, that can fulfill the full set of
aforementioned tests and physical constraints, is not an easy task. In fact, the corresponding
Einstein and field equations are extremely nonlinear, due to the Lagrangian associated with
the tachyon field. Despite these difficulties, a great effort has been made in \cite{GADRR} to
provide an interesting tachyonic thick braneworld supported by a tachyon scalar field coupled
to gravity with a bulk cosmological constant that yields certain phenomenological aspects of our
4D Universe. This model possesses several appealing properties: a) it contains an expanding metric induced on the 3--brane which arises as an exact
solution to the 5D field equations; b) the field configuration is completely {\it regular and stable} under small perturbations \cite{German:2015cna}; c) The de Sitter 3--brane describes the inflationary epochs of our universe; d) the braneworld is {\it asymptotically flat}, despite the presence of a negative 5D cosmological constant (usually braneworlds are asymptotically dS or AdS); e) it contains a graviton spectrum with a single massless bound state that accounts for 4D gravity localized
on the brane; f) it has a mass gap that makes the 5D corrections to Newton's and Coulomb's laws decay exponentially; g) finally, gauge, scalar and fermion fields were shown to be localized on this braneworld.

Thus, in this work we show that this tachyonic braneworld model allows for a {\it more general}
nontrivial solution with no bulk cosmological constant,  that
preserves all the above mentioned remarkable properties with a less
amount of parameters, constituting an important contribution to the
construction of a realistic cosmological braneworld model.

The paper is organized as follows: In Sect. 2 the tachyonic thick braneworld model with the 5D cosmological constant is briefly reviewed. In Sect. 3,   we obtain a second exact solution {\it without} the bulk cosmological constant, also analyzing its underlying parameters and some relevant derived quantities (the tachyonic self--interaction potential, the effective 4D Planck mass, and the curvature scalar), hence comparing them with their expressions
when the bulk cosmological constant was present. Finally, a brief discussion of the new solution for our tachyonic braneworld model is presented and elucidated as a set of
conclusions in Sect. 4.

\section{Review of the tachyonic de Sitter thick braneworld }

The  5D action for a thick braneworld model that is produced by gravity and a bulk tachyonic scalar field
in the presence of a 5D cosmological constant reads \cite{GADRR}
\begin{equation}
S = \int d^5 x \sqrt{-g} \left(\frac{1}{2\kappa_5^2} R - \Lambda_5 -
V(T)\sqrt{1+g^{AB}\partial_{A} T\partial_{B} T} \right),
\label{action}
\end{equation}
where $R$ is the 5D scalar curvature, $\Lambda_{5}$ is the bulk
cosmological constant, the tachyon field $T$ represents the matter
in the 5D bulk, $V(T)$ denotes its self--interaction potential,
$\kappa_5^2=8\pi G_5$ with $G_5$ being the
5D Newton constant, and  $A, B =0,1,2,3,5.$
Besides, a 5D metric ansatz with an induced 3--brane of  FLRW
background type is be taken to be
\begin{eqnarray}
\label{metricw}
ds^2 = e^{2f(w)} \left[- d t^2 + a^2(t) \left(d x^2 + d y^2 + d
z^2 \right)+ d w^2 \right],
\end{eqnarray}
where $\text{e}^{2f(w)}$ and $a(t)$ are the warp factor and the
scale factor of the brane, respectively, and $w$ represents the extra--dimensional coordinate.
The corresponding Einstein's equations for this model read
\begin{eqnarray}
\label{EinsteinEq_5d}
G_{AB} = - \kappa_5^2 ~\Lambda_5 g_{AB} + \kappa_5^2 ~T_{AB},
\end{eqnarray}
where $T_{AB}$ is the energy--momentum tensor for the bulk tachyonic
scalar field,  described by
\begin{equation}
T_{AB} =  \left[ - g_{AB} \, V(T) \sqrt{1 +
(\nabla T)^2} + \frac {V(T)}{\sqrt{1+ (\nabla T)^2}} \,
\partial_{A} T \, \partial_{B} T \right].
\end{equation}
The matter field equation corresponding to the action (\ref{action})
is expressed in the following form:
\begin{equation}
\Box T-\frac{\nabla_C\nabla_D T\,\, \nabla^{C} T\,\, \nabla^{D}
T}{1+(\nabla T)^2}= \frac{1}{V} \frac{\partial V(T) }{\partial T}.
\label{fieldequ1}
\end{equation}

By using the metric ansatz (\ref{metricw}) and the fact that the tachyon field depends only on the extra coordinate in (\ref{EinsteinEq_5d}) and (\ref{fieldequ1}) (i. e. $T=T(w)$), it is straightforward to write a set of nonlinear coupled field equations, which have the following complete solution:
\begin{equation}
a(t)=e^{H\,t}, \qquad  \qquad f(w)=\frac{1}{2}\ln\left[s\,{\rm
sech}\left(\,H\,(2w+c)\right)\right], \label{scalewarpfactors}
\end{equation}
%whereas the tachyon scalar field is given by
\begin{eqnarray}
T(w) &=& \pm\sqrt{\frac{-3}{2\,\kappa_5^2\,\Lambda_5}}\
\mbox{arctanh}\left(\frac{\sinh\left[\frac{H\,\left(2w+c\right)}{2}\right]}
{\sqrt{\cosh\left[\,H\,(2w+c)\right]}}\right), \label{Tw}
\end{eqnarray}
%with its self-interacting potential
\begin{eqnarray}
V(T) &=& - \Lambda_5\
\mbox{sech}\left(\sqrt{-\frac{2}{3}\kappa_5^2\,\Lambda_5}\ T\right)
\sqrt{6\ \mbox{sech}^2\left(\sqrt{-\frac{2}{3}\kappa_5^2\,\Lambda_5}\ T\right)-1}\nonumber \\
&=& - \Lambda_5\ \sqrt{\left(1 +
\mbox{sech}\left[\,H\,(2w+c)\right]\right) \left(1+\frac{3}{2}\
\mbox{sech}\left[\,H\,(2w+c)\right]\right)}, \label{VT1}
\end{eqnarray}
where $H$, $c$ and $s>0$ are arbitrary constants. By consistency of
the above presented field equations, the constant $s$ must be
\begin{equation}
\label{spar}
s=-\frac{6H^2}{\kappa_5^2\,\Lambda_5},
\end{equation}
where the arbitrary 5D cosmological constant is negative,
$\Lambda_5<0$, to ensure the real nature of the tachyonic field and its
potential. It is important to note that we have an explicit
expression for the self-interaction potential in terms of the
tachyon field. This potential has a maximum at the
position of the brane. Since the tachyonic field has a bounded domain in
order to be real, it then leads to a real and bounded potential as
well.

By looking at the structure of Eq. (\ref{spar}) one can see that the field
configuration has a limiting case when the bulk cosmological constant
vanishes $\Lambda_{5} \rightarrow 0$  with the same rapidity as
$H^2  \rightarrow 0$ in such a way that $s$ remains finite. In this two-fold
limit, the tachyon field becomes linear $T =\pm \frac{\sqrt{s}}{4}(2w+c)$,
the self-interaction potential vanishes $V(T) = 0$, whereas the scale and warp
factor become a constant that can be further ignored through a rescaling of
the metric coordinates, leading to a flat 5D spacetime \cite{GADRR}.

The corresponding 5D curvature scalar reads
\begin{equation}
R=-\frac{14}{3}\kappa_5^2\,\Lambda_5\,\mbox{sech}\left[\,H\,(2w+c)\right],
\label{R}
\end{equation}
which is positive definite and asymptotically flat along the
fifth dimension. This assertion is easy to see by observing the
action (\ref{action}) along with the asymptotic behavior of the
potential (\ref{VT1}) which cancels the value of the 5D
cosmological constant, while $T$ remains constant at infinity.

An important point that should be stressed is that this 5D spacetime is completely
free of naked singularities, that usually arise when the mass spectrum of
Kaluza-Klein tensorial fluctuations display a mass gap, giving rise to an
important feature of the graviton mass spectrum from the phenomenological point
of view (see Refs. \cite{gravity1,gravity2,gravity3}
% and \cite{thickbraneworlds}
for further details).

A detailed analysis of the stability of  the tensorial metric fluctuations of the
tachyonic braneworld was performed in \cite{GADRR}, showing that 4D gravity is
consistently localized on it. Moreover, the corrections to the Newton's law, due to the
5D massive fluctuations, were computed in the thin brane limit and shown to be
exponentially suppressed. On the other hand, the stability of the model under
small fluctuations of the tachyon field and the scalar metric modes was successfully
achieved with the aid of an auxiliary Sturm--Liouville eigenvalue problem for any
value of the bulk cosmological constant in \cite{German:2015cna}.

\section{The thick brane with $\Lambda_5=0 $ model and its solution}\label{cap:New solution}

The action (\ref{action}) that describes our tachyonic braneworld model presents a second exact solution
that can be obtained from the Einstein's equations when the bulk cosmological constant
vanishes
\begin{equation}
 G_{AB} =  \kappa_5^2 ~T_{AB}.
\label{einequ}
\end{equation}
By making use of the metric ansatz (\ref{metricw}), the Einstein tensor components
read
\begin{eqnarray}
G_{00} &=& 3\, \frac{\dot a^2}{a^2} - 3\, \left( f^{''} +  f^{'2}
\right),
\label{eqeintach1}\nonumber \\
G_{\alpha\alpha} &=& - 2\, \ddot aa - \dot a^2 + 3\, a^2  \left(
f^{''} +  f^{'2} \right) \label{eqeintach2},
\nonumber\\
G_{5 5} &=& -3\,\left( \frac{\ddot a}{a} + \frac{\dot a^2}{a^2}
\right) + 6f^{'2}, \label{eqeintach3}
\end{eqnarray}
where ${``\prime"}$ and ${``\cdotp"}$ are derivatives with
respect to the extra dimension and time, respectively, whereas the index
$\alpha$ labels the spatial dimensions $x,$ $y$ and $z$.

The main goal of this work is to construct a new relevant solution
in the context of the above mentioned tachyonic thick braneworld
with a vanishing 5D cosmological constant and analyze whether the
appealing properties of the brane field configuration
(\ref{scalewarpfactors})--(\ref{VT1}) are still present in the new
solution.

By taking into account the fact that the tachyon field depends
only on the fifth dimension, $T(w)$, Eq. (\ref{fieldequ1}) adopts the form
\begin{equation}
T^{''}-f^{'}T^{'}+4f^{'}T^{'}(1+e^{-2 f}T^{'2})=(e^{2
f}+T^{'2})\frac{\partial_{T}V(T) }{V(T)}.
\label{fieldequ}
\end{equation}
At the same time the Einstein's equations (\ref{einequ}) can be
rewritten as
\begin{eqnarray}
f^{''}- f^{'2} + \frac{\ddot a}{a} &=& -
\kappa_5^2\frac{V(T)T^{'2}}{3\sqrt{1 +  e^{-2 f}\,T^{'2}}} ,
\label{einsteinequ} \\
f^{'2} - \frac{1}{2} \left(\frac{\ddot a}{a}+\frac{\dot a^2}{a^2}
\right)&=& - \kappa_5^2\frac{e^{2 f}\,V(T)}{6\sqrt{1 + e^{-2
f}\,T^{'2}}} .
\label{restriccion}
\end{eqnarray}
By demanding consistency of this system of equations, the 3--brane solution
corresponds to a 4D de Sitter cosmological background defined by
\begin{equation}
a(t)= e^{H\,(t-t_0)}, \label{scalefactor}
\end{equation}
however, the constant factor $a_0=1/e^{H\,t_{0}}$ can be absorbed by a rescaling
of the spatial coordinates. Thus, this result in a braneworld model in which the induced
metric on the 3--brane is described by a $dS_4$ geometry.

By further taking into account the form of the scale factor $a(t)= e^{H\,t}$,
we can recast (\ref{einsteinequ}) and (\ref{restriccion}) as follows
\begin{eqnarray}
f^{''} - f^{'2} + H^2 &=& - \kappa_5^2\frac{V(T)T^{'2}}{3\sqrt{1 +
e^{-2 f}T^{'2}}},
\label{Ee1z} \\
f^{'2} - H^2 &=& - \kappa_5^2\frac{e^{2 f}V(T)}{6\sqrt{1 + e^{-2
f}T^{'2}}}. \label{Ee2z}
\end{eqnarray}
Now it is straightforward to obtain separate equations for the derivative of the scalar field $T$
and the self-interaction potential $V(T)$ from Eqs. (\ref{Ee1z}) and (\ref{Ee2z}), leading to:
\begin{eqnarray}
T'&=&\pm e^f\sqrt{\frac{f''-f'^2+H^2}{2\left(f'^2-H^2\right)}},
\label{Tprime}\\
V(T)&=& \pm \frac{3}{\kappa_5^2}\,e^{-2f}\,\sqrt{2\left(f''+
f'^2-H^2\right)\left(f'^2 -H^2\right)}. \label{V}
\end{eqnarray}
Therefore, by determining the nature of the scaling and warp factors,
the self-interaction potential $ V(T) $ and the derivative of the tachyon
field $T'$ are completely determined. However, care must be taken, since the tachyon scalar field must be real and have a form that ensures
the localization of 4D gravity, whereas the self-interaction potential must be
real and well defined in the sense that ensures the stability of the whole
braneworld field configuration.
These restrictions are very demanding, several warp factors with
``convenient" localizing behavior lead to a complex tachyon field $T$.
On the other hand, ensuring a real and stable tachyon potential often
yields either a complex tachyon field $T$ or a warp factor that does not
enable the localization of 4D gravity.
Hence, we propose the most general form for the warp factor as follows
\begin{equation}
\label{fwg}
f(w)=-n\ln\left[\frac{\cosh\left[\,H\,(\frac{w}{n}+c)\right]}{s}\right],
\end{equation}
where $H$, $c$, $s$, and $n$ are constants. Here we should demand $s>0$ and $n>0$ in order to have a warp factor that ensures the localization of 4D gravity and other matter fields. It is straightforward to realize that this warp factor is a solution of the Einstein and field equations if the tachyon scalar field adopts the {\it general} form
\begin{equation}\label{Twg}
T(w) = \pm i\frac{\sqrt{\frac{n}{2 (1-n)}}\,\, s^n \,\,
_2F_1\left(\frac{1}{2},\frac{1-n}{2};\frac{3-n}{2};\cosh ^2\left[H
\left(\frac{w}{n}+c\right)\right]\right)}{H \cosh ^{n-1}\left[H
\left(\frac{w}{n}+c\right)\right]} + k, \quad    n\neq 1,
\end{equation}
where $k$ denotes a pure imaginary number. The tachyon scalar field is purely real within the domain $0<n<1$ andm under a suitable choice of the parameter $k$ (for non--integer values $n>1$ the tachyon field is purely imaginary), and the tachyon potential is given by the following expression
\begin{equation}\label{VTg}
V(w)=\frac{3 H^2 \sqrt{\frac{2 (n+1)}{n}} }{k_5^2 s^{2
n}}\text{sech}^{2(1-n)}\left[H \left(\frac{w}{n}+c\right)\right].
\end{equation}

It is convenient to have at hand a real tachyon scalar field that allow us to
explicitly write the self-interaction potential $V(w)$ in terms of the scalar field $T$. In order to
satisfy both conditions we consider the particular case when $n=\frac{1}{2}$. For this specific
value, the warp factor, the tachyon scalar field and the self-interaction potential read
\begin{eqnarray}
f(w)
&=&-\frac{1}{2}\ln\left[\frac{\cosh\left[\,H\,(2w+c)\right]}{s}\right],
\label{fw}\\
T(w) &=& \pm \frac{1}{\sqrt{2}b}\,i\,\text{EllipticF}\left( i H
\left(
w+\frac{c}{2}\right),2\right), \label{Tw2}\\
V(T) &=& \frac{3 \sqrt{6}\, b^2}{k_5^2 }\,\text{sec}\left[2\text{JacobiAmplitude}\left(i \sqrt{2} b T,2\right)\right]\nonumber \\
&=& \frac{3 \sqrt{6}\, b^2}{k_5^2 }\,\text{sech} \left(H \left(2
w+c\right)\right), \label{VT}
\end{eqnarray}
where the $\text{JacobiAmplitude}(u,m)$ gives the amplitude $\text{am}(u | m)$ for Jacobi elliptic functions, i. e., it is the inverse of the elliptic integral of the first kind.
%If $u=F(\phi | m)$, then $\phi = \text{am}(u | m)$.
In the last two equations we set
\begin{equation}
s=\frac{H^2}{b^2}.
\end{equation}
where $b$ is an arbitrary constant and the scale factor remains the same.
%%%%%%%%%%%%%%%%%%FIGURA%%%%%%%%%%%%%%%%%%%%%%%%%%%%%%%%%%%%%%%%%%%
\begin{figure}[htb]
\begin{center}
\includegraphics[width=7cm]{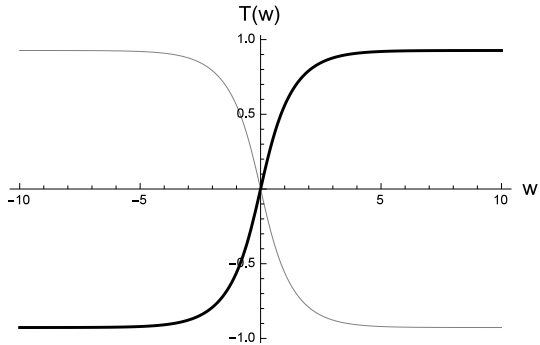}
\end{center}\vskip -5mm
\caption{The profile of the tachyonic scalar field $T$. The thin
line represents the positive branch of the field and the thick line
is associated to the negative  branch of the tachyon.  Here we have
set $n=1/2$, $c=0,$ $H=1$, $2\kappa_{5}^2=1$ and $s=1$ for
simplicity.} \label{fig_T}
\end{figure}
%%%%%%%%%%%%%%%%%%%%%%%%%%%%%%%%%%%%%%%%%%%%%%%%%%%%%%%%%%%

%%%%%%%%%%%%%%%%%%%%FIGURA%%%%%%%%%%%%%%%%%%%%%%%%%%%%%%%%%%%%%%%%%%%%%%%%%%%%%%
\begin{figure}[htb]
\begin{center}
\includegraphics[width=7cm]{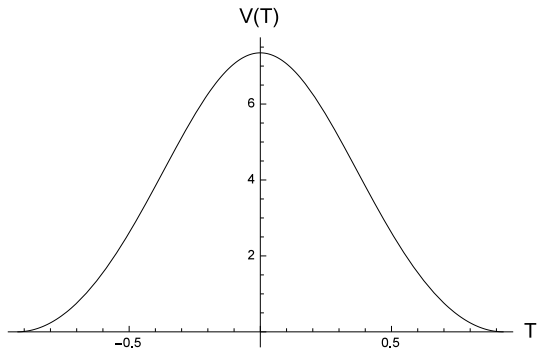}
\end{center}\vskip -5mm
\caption{The shape of the self--interaction potential of the
tachyonic scalar field $V(T)$. We set  $n=1/2$, $c=0,$ $H=1$,
$2\kappa_{5}^2=1$ and $s=1$ for simplicity.} \label{fig_VT}
\end{figure}
%%%%%%%%%%%%%%%%%%%%%%%%%%%%%%%%%%%%%%%%%%%%%%%%%%%%%%%%%%%

Therefore, the set of equations (\ref{scalefactor}), (\ref{fw}), (\ref{Tw2}), and (\ref{VT}) provide
a new relevant tachyonic thick braneworld configuration. By comparing with the previous solution
(\ref{scalewarpfactors}), (\ref{Tw}) and (\ref{VT1}), the new warp factor (\ref{fwg}) also has a decaying and
vanishing asymptotic behavior for $0<n<1$ and coincides with (\ref{scalewarpfactors}) when $n=1/2$. Similarly, the new tachyon scalar field (\ref{Tw2}) is real and possesses a kink-- or antikink--like profile as shown in Fig. 1, but with the positive and negative branches opposite to those presented by the field (\ref{Tw}).  On the other hand, the potential (\ref{VT}) also has a maximum at the position of the brane, it is positive definite and bounded (from below and above), but unlike (\ref{VT1}), it vanishes as $w$ tends to infinity or, equivalently, as $T$ approaches it asymptotic value (see Fig. 2), emphasizing the fact that the braneworld is asymptotically flat.

It is important to recall that the new solution presented in this work cannot be recovered from the solution (\ref{scalewarpfactors})-(\ref{VT1}) with the aid of a two-fold  limit. On the other hand, it can be noticed as well that
when $\Lambda_5=0$, the tachyon scalar field (\ref{Tw}) diverges, while the potential (\ref{VT1}) vanishes,
indicating that this is not a physical limit.

By substituting Eqs. (\ref{Twg}) and (\ref{VTg}) into Eqs. (\ref{Ee1z}) and (\ref{Ee2z}), we can see that there is no restriction on the $s$ parameter coming from the field equations of the system. This contrasts  with the first solution generated by the warp factor (\ref{scalewarpfactors}) with $\Lambda_5 \neq 0$,  where the field equations demand the relation $s=-\frac{6H^2}{\kappa_5^2\,\Lambda_5} \label{s}$.

However, when we compute the effective 4D Planck mass $M_{pl}$ for the special value $n=\frac{1}{2}$ with no bulk cosmological constant $\Lambda_{5}=0$, then we obtain $s=H^2/b^2$, where $b$ is an arbitrary constant, yielding
\begin{equation}\label{Jm1}
M_{\text{Pl}}^{2} \sim M_*^3 \frac{\sqrt{\frac{2}{\pi }}  \Gamma
\left(\frac{3}{4}\right)^2 H^2}{b^3}.
\end{equation}
\noindent This expression shows the relationship  between the Planck mass in
5D and 4D, adjusted with the $b$ parameter.  Comparing to the braneworld
model presented in \cite{GADRR}, the effective Planck mass in 4D is related to the 5D one by
\begin{equation}
\label{Jm2}
M_{\text{Pl}}^{2} \sim - M_*^3 \frac{\sqrt{\frac{2}{\pi }}\Gamma
\left(\frac{3}{4}\right)^2 H^2}{\Lambda_5^3},
\end{equation}
which diverges when taking the single limit $\Lambda_5 \rightarrow 0$. However, one can still perform a two-fold limit, leading to a finite value of the 4D Planck mass $M_{\text{Pl}}^{2}$.

%This important fact shows that the solution here presented is mathematically different  and possess a different physical nature.

Finally, computing the 5D curvature scalar for our solution (\ref{fwg}) yields
\begin{equation}
R=\frac{4 H^2 (3 n+2)}{n\,s^{2 n} } \text{sech} ^{2 (1-n)}\left[H
\,\left(\frac{w}{n}+c\right)\right],
\label{R5n}
\end{equation}
in agreement with the asymptotic form dictated by the
self-interaction  potential (\ref{VT}). For the special value $n=1/2$, it reads
\begin{equation}
R=28\,b^2\,\mbox{sech}\left[\,H\,(2w+c)\right]. \label{R5}
\end{equation}

This 5D invariant is positive definite and asymptotically vanishes, yielding an asymptotically
5D Minkowski spacetime as the solution (\ref{scalewarpfactors})-(\ref{VT1}) does.
However, unlike the curvature scalar corresponding to the braneworld \cite{GADRR}, this 5D invariant is
obtained within a simpler theory, since the original action (\ref{action}) has one less input parameter
($\Lambda_5 =0$).

\section{Discussion and conclusions}

We presented a new complete solution for the braneworld model (\ref{action}), generated
by a tachyonic scalar field minimally coupled to gravity, with no bulk cosmological constant,
and a 5D warped metric ansatz (\ref{metricw}), with respect to which the 3--brane metric defines a FLRW  geometry.
The resulting spacetime is regular and stable  along the whole fifth dimension for certain values of
the $n$ parameter. Analytic expressions were derived for the physically meaningful warp factor (\ref{fwg}), the
tachyon scalar field (\ref{Twg}), the self-interaction potential (\ref{VTg}) and the curvature scalar
(\ref{R5n}), for the significant values of the $n$ parameter (for $0<n<1$).

On the one hand, the profile of the warp factor (\ref{fwg}) and the 5D Ricci scalar (\ref{R5n}) show a regular structure within the domain $n\in(0,1)$. These geometrical quantities show that 4D gravity, as well as other Standard Model matter fields, can be localized in the tachyonic de Sitter braneworld model, whenever  $0<n<1$. On the other hand, the tachyon scalar field (\ref{Twg}) and its self-interaction potential (\ref{VTg}) are real only for the aforementioned domain $0<n<1$. Moreover, the potential (\ref{VTg}) leads to a completely {\it stable} braneworld configuration, in full compliance to the treatment presented in \cite{German:2015cna}. In fact, it has a maximum for precisely this range of the $n$ parameter.

It is worth noticing as well a physically consistent picture at the level of the Einstein's equations: the geometrical left hand side localizes the known 4D interactions of our world and renders a regular, asymptotically flat braneworld for $0<n<1$, whereas the matter right hand side is real and makes the whole field configuration stable, for the same domain $0<n<1$ within our setup.

It is important to note that the 5D spacetime, for the special value $n=\frac{1}{2}$, supports the same geometrical configuration as the solution presented in \cite{GADRR}. However, it has one parameter less involved in the initial braneworld action, since $\Lambda_5=0$. This fact yields the localization properties of the model, with $n=\frac{1}{2}$ remaining  the same, since the form of the warp factor determines the character of the 5D graviton spectrum of KK massive modes (which can be gapless or possess a mass gap, for instance) and its localizing properties.

Thus, the braneworld solution (\ref{fw}) presented in Section \ref{cap:New solution} preserves all the aforementioned appealing properties that we look for in such a model for being realistic. In fact, it is completely regular and stable under small metric and tachyon field perturbations, it localizes 4D gravity and from the low-energy point of view recovers the Newton's law \cite{GADRR}  and the Coulomb's law \cite{coulombcorrec} in the thin brane limit with small, exponentially suppressed corrections that come  from the extra dimension. Moreover, the structure of the warp factor given in Eq. (\ref{fw}) for this model makes possible the localization of massive (massless) scalar fields with spin-$0$ \cite{AARRH}, gauge boson fields with spin--{1} \cite{ahagauge}, and spin--$1/2$ fermions  \cite{coulombcorrec}.

It seems that the localization of these matter fields can be achieved for any value of the $n$ parameter within the interval $0<n<1$. However, it is not straightforward to invert the relation we obtain upon integration of (\ref{Tprime}) in order to get an explicit expression for the tachyonic scalar field in terms of the fifth dimension and an analytic expression for the self-interaction potential in terms of the tachyon field for a value $n\ne1/2$, representing a major technical difficulty that we are currently trying to address.

Finally, one might use the information entropy on thick branes \cite{Correa:2016pgr,Correa:2015vka}, to determine a more precise bound on the $n$ parameter  and to derive refinements of the model here presented.

\section*{Acknowledgments}

NBC is grateful to FIE-UMSNH for the designation of a working space and to Juan Herrera for technical support. The work of A.H.-A. was completed at the Aspen Center for Physics, which is supported by National Science Foundation grant PHY-1607611 and a Simons Foundation grant as well. He expresses his gratitude to the ACP for providing an inspiring and encouraging atmosphere for conducting part of this research. RCF and AHA acknowledge a VIEP-BUAP grant. RCF, NBC, AHA, and RRML thank SNI for support. RdR~is grateful to CNPq (Grant No. 303293/2015-2),
and to FAPESP (Grants No.~2015/10270-0 and No.  2017/18897-8), for partial financial support.

\end{document}